\documentclass[showpacs,aps,pra,twocolumn]{revtex4}
\usepackage{graphicx,epsfig}
\usepackage{times}
\usepackage{amsmath,amssymb,wasysym}
\usepackage{color}
\usepackage[normalem]{ulem} 
 
\usepackage{braket}

\DeclareMathOperator*{\argmin}{arg\,min}

\newlength\figurewidth
\newcommand{\rem}[1]{}

\newcommand{\R}{\mathbb{R}}

\newcommand{\mean}[1]{\langle#1\rangle}

\begin{document}

\title{Determination of the full statistics of quantum observables using the maximum entropy method}
  \author{Boris Gulyak}
  \affiliation{Institut f{\"u}r Physik, Otto-von-Guericke-Universit{\"a}t Magdeburg, Postfach 4120, D-39016 Magdeburg, Germany}
  \affiliation{Institut f{\"u}r Analysis und Numerik, Otto-von-Guericke-Universit{\"a}t Magdeburg, Postfach 4120, D-39016 Magdeburg, Germany}
  \author{Boris Melcher}
  \affiliation{Institut f{\"u}r Physik, Otto-von-Guericke-Universit{\"a}t Magdeburg, Postfach 4120, D-39016 Magdeburg, Germany}
  \author{Jan Wiersig}
\email{jan.wiersig@ovgu.de}
  \affiliation{Institut f{\"u}r Physik, Otto-von-Guericke-Universit{\"a}t Magdeburg, Postfach 4120, D-39016 Magdeburg, Germany}
\date{\today}
\begin{abstract}
Numerical methods for the description of nonequilibrium many-particle quantum systems such as equation of motion techniques often cannot compute the full statistics of observables but only moments of it, such as mean, variance and higher-order moments. We employ here the maximum entropy method to numerically construct unbiased statistics based on the knowledge of moments. We verify the feasibility of the proposed method by numerical simulation of a simple birth-death model for quantum-dot-microcavity lasers, where the full photon and carrier statistics are available for comparison. We show that not only the constructed statistics but also the computed entropy and the Lagrange multipliers, which appear here as a byproduct, provide valuable insight into the physics of the considered system. For example, the entropy reveals that, in contrast to common wisdom, the photon statistics of the microcavity laser above threshold is better described by a Gaussian distribution than by a Poisson distribution. 
Our approach is general and can be applied to many other systems emerging in physics and related fields.
\end{abstract}
\maketitle

\section{Introduction}
The study of semiconductor quantum dot (QD) microcavities has been a subject of considerable attention due to their high potential, e.g., for single-photon sources~\cite{Michler2000}, sources of entangled photon pairs~\cite{BSP00,HZS17}, and ultra-low threshold lasing~\cite{WGJ09}. The resonator of such a novel laser is given by an optical microcavity~\cite{Vahala03} with small mode volumes and small cavity losses which can be used to increase the spontaneous emission factor of the laser leading to the so-called thresholdless laser~\cite{RiceCarmichael94,OKW17}.
 
From the theoretical point of view, QD-microcavities are driven-dissipative quantum many-particle systems. A straightforward way to describe the dynamics of such a nonequilibrium system is to numerically solve the von-Neumann-Lindblad equation for the reduced density operator~\cite{Carmichael98}. However, this is only feasible for highly symmetric~\cite{GR16} or sufficiently small systems, such as a single-QD laser~\cite{GFG11}. 
For other systems, it is more appropriate to derive the equations of motion (EoM) directly for the quantities of interest~\cite{LFW14}, such as expectation values of photon numbers~$\mean{n}$ or higher moments~$\mean{n^2}$ etc. In this way, however, an infinite hierarchy of coupled differential equations is unfolded which has to be truncated in one way or the other. EoM techniques have been used successfully to realize microscopic descriptions of quantum systems, and are a way to systematically incorporate many-particle correlations into the description of exciton dynamics in semiconductor quantum wells~\cite{Hoyer2003}, ultracold Bose gases~\cite{KB02}, spin dynamics~\cite{KP08}, photoluminescence from quantum wells~\cite{Jahnke1998} and QDs~\cite{BGWJ06,Feldtmann06,FGJ13}, resonance fluorescence from quantum wells~\cite{PQE}, cavity phonons~\cite{KCB12}, cavity-quantum-electrodynamics~\cite{CRC10}, QD-microcavity lasers~\cite{GWLJ06,LHA13,KLH15,FFL16}, and superradiant emission from QD-microcavity systems~\cite{LFW15,JGA16}. A freely available source code for EoM schemes is provided in~\cite{FLW17}. 

The problem with EoM techniques is that they do not provide the full statistics but only moments of an order limited by the truncation level. For instance, the approach in~\cite{GWLJ06} for QD-microcavity lasers gives $\mean{n}$ and $\mean{n^2}$ which is enough to determine the intensity and the second-order autocorrelation function 
\begin{align}
	g^{(2)}(0)=\frac{\langle n^2\rangle-\langle n\rangle}{\langle n\rangle^2} \label{eq:g2}
\end{align} 
of the emitted light. But it is not enough to determine higher-order moments of the photon statistics (the photon number distribution $p_n$) needed for the computation of higher-order autocorrelation functions $g^{(3)}(0)$ and $g^{(4)}(0)$ which can be also measured nowadays~\cite{WGJ09,AVB09}. Moreover, in some cases more details of the photon statistics $p_n$ are required for a clear interpretation of, e.g., single-photon sources based on a few QDs~\cite{GJC15} and mode competition in two-mode microcavity lasers~\cite{LHA13}. Recent experiments allow to determine the full photon statistics $p_n$ of a single-mode light source using a transition-edge sensor~\cite{SHL18}.

The maximum-entropy method (MEM) is a widely-used procedure to estimate a probability distribution by maximizing the Shannon information for given constraints~\cite{Wu46}. Originally, the method was introduced to determine the density operator and the entropy in equilibrium statistical mechanics~\cite{Jaynes57}. The MEM has been extended to nonequilibrium situations within the concept of observation levels which includes all known moments into the construction of the density operator, see, e.g.,~\cite{FickSauermann90}. 

The aim of this paper is to exploit the MEM for an extension of the range of applications of EoM techniques. The moments computed by an EoM technique can be used as constraints to determine approximately the full statistics of a given quantum mechanical system in an unbiased way. The MEM not only provides the full statistics but also the entropy and the Lagrange multipliers. We demonstrate that these quantities can be used to characterize QD-microcavity systems. 
As a benchmark model we consider the birth-death model by Rice and Carmichael~\cite{RiceCarmichael94} of a microcavity laser with discrete light emitters. This kind of master equation is phenomenological in nature but has the advantage that the full statistics can be computed for comparison. 

The outline of the paper is as follows. In Sec.~\ref{sec:mem} we briefly review the MEM. The birth-death model is explained in Sec.~\ref{sec:bdm}. In Sec.~\ref{sec:imp} we discuss in detail the implementation of our method. Numerical results are presented in Sec.~\ref{sec:num}. Conclusions are given in Sec.~\ref{sec:con}.

\section{Maximum-entropy method}
\label{sec:mem}
The method of entropy maximization has its roots in a Bayesian, information-theoretical view of statistical physics developed by E. T. Jaynes~\cite{Jaynes57}. Due to uncertainty, randomness or exorbitant complexity of physical problems, one often has only few  data. The aim of this method is a reasonable statistical inference from available knowledge, like values of moments, to lacking system probability distributions. In general, there are many admissible extensions out of the known data, so the point is to select a guideline that is justifiable by objective reasoning. A good attempt is made by the MEM where the inference is done by {choosing the most unbiased probability distribution that satisfies the conditions of known a-priori information}.

 The bias can be quantified by an information measure named entropy or {Shannon information}. For a {discrete probability distribution} $p=(p_n)_{n=0}^\infty$, it is defined by~\cite{Jaynes57}
 \begin{align}
	{S(p)=-\sum_n p_n \ln p_n}.  \label{eq:h} 
\end{align}
 This is up to the Boltzmann constant $k_{\text{B}}$ exactly the thermodynamic Gibbs entropy, whereby $p_n$ plays the role of a microstate's probability. Mathematically,  $S$ is a concave, positive functional on the space of probability distributions.
 
 Actually, entropy can be seen as the measure of uncertainty  of the statistical distribution to predict the measurement outcome. For example, a distribution $(1,0,0,\dots, 0)$, which predicts the outcome with certainty, has a minimal entropy of zero. On the other hand, the maximum entropy distribution with $n$ possible different values is the equiprobable $(\frac1n,\frac1n,\dots,\frac1n)$ that exhibits the smallest forecasting power. The other distributions interpolate between these two limits: with smaller entropy, the concentration of probability increases. 
 
Furthermore, the entropy is {invariant under reordering} of the probabilities, thus the Shannon information is unable to distinguish between unimodal and multimodal distributions, because it is always possible to sort the probabilities into a unimodular distribution. Moreover, the concentration quantified by the entropy is non-local in the sense that the probabilities are interchangeable, thus do not have neighbors from the entropy's point of view. 

Having understood the entropy as a measure of uninformativeness, we can recall the {maximum entropy principle} (MEP) for statistical inferences~\cite{Jaynes57}: {When an inference is made on the basis of incomplete information, it should be drawn from the probability distribution that maximizes the entropy subject to the constraints on the distribution.}
 We call such a distribution a maximum entropy distribution (MED). It is important to note, that in general the existence of a MED cannot   be guaranteed~\cite{MP84}, some necessary conditions will be given in Sec.~\ref{subsec:bounds}. 

Yet, the most convenient statistical information is the knowledge of expectation values, like {moments}. For illustrating and making use of the MEP, let the first $k$ {moments} be  $\langle n\rangle=\mu_1, \langle n^2\rangle=\mu_2, \dots, \langle n^k\rangle=\mu_k$ as known partial information as basis for inference. We formulate these as constraints for a distribution $(p_n)_{n=0}^\infty$,
\begin{align}
	\big\langle n^i\big\rangle=\sum_n n^{i}p_n=\mu_i,\quad i=1,\dots,k. \label{eq:nb} 
\end{align} 
To obtain the MED with the MEP, it is sufficient to solve the following concave optimization problem:
\begin{align*}
	\text{maximize }\ S(p), \  \text{subject to:}\quad \big\langle n^i\big\rangle=\mu_i,\quad i=1,\dots,k.
\end{align*}
 Using the ordinary procedure of  maximization under constraints with Lagrange multipliers $(\lambda_i)_{i=1}^k$ one uniquely~\cite{Einbu77} obtains the so called $k^\text{th}$-order MED~\cite{Jaynes57}
\begin{align}
	p^\text{\text{MED}}_{n}=\frac{1}{Z(\lambda)}\exp\left(-\sum_{i=1}^k\lambda_i n^i\right) \ ,  \label{eq:mev}
\end{align}
 with $(\lambda_i)_{i=1}^k$ determined selfconsistently by $(\mu_i)_{i=1}^k$ in the way that the constraints \eqref{eq:nb} are fulfilled. The normalization constant is a partition function like in statistical physics
\begin{align*} 
	Z(\lambda)=\sum_n \exp\left(-\sum_{i=1}^k\lambda_i n^i\right) \ .
\end{align*} 

For such MED to exist on $\mathbb N_0=\{0,1,\dots,\infty\}$, the first necessary condition is a positive last Lagrange multiplier $\lambda_k \geq 0$. Otherwise,  the MED will grow to infinity for $n\to\infty$, and thus will not be normalizable. This circumstance is related to the question of positivity of absolute temperature for canonical ensembles, thus if the configuration space is $\mathbb N_0$, then $T>0$. On the contrary, for a finite configuration space also negative absolute temperatures are possible~\cite{amann_mueller11}.

\section{Birth-death model}
\label{sec:bdm}
In this article, we test the MEM by considering the birth-death model of Rice and Carmichael~\cite{RiceCarmichael94}. We would like to stress that our method is not limited to this model nor to quantum optics. Moreover, many later discussions are pure mathematical or numerical, and so they can be used for other models emerging in physics, biology, and economy.

The birth-death model is a  stochastic model and serves as a description of the  quantized single-mode light field coupled via single-electron excitations in atomic or QD systems. Instead of using the complete density matrix $\rho$ (see, e.g.,~\cite{SL67}) whose dynamics is described by the von-Neumann-Lindblad equation, it characterizes the system only by the diagonal elements of $\rho$: the probability  $p_{n,N} $ of states with $n$ photons in the laser mode and $N$ atoms or carriers in the upper level. This reduces the dimensionality and allows to compute the full statistics. 

There are three approaches to derive the birth-death master equation. The first is the approximation of the von-Neumann-Lindblad equation by adiabatically eliminating the off-diagonal matrix elements of $\rho$ to get a closed system for $p_{n,N} $~\cite{gartner2016laser}. The second is a mathematical extrapolation out of rate equations by replacing the moments by weighted probabilities~\cite{Carmichael98}. The third and most direct way is to let all relevant processes act as phenomenological transition rates between probabilities, which we use in the following discussion. Thus, the time-derivative of $p_{n,N} $ is determined by neighboring states $p_{n\pm 1,N\mp1} $, $p_{n,N\mp1} $, $p_{n\pm 1,N} $ and $p_{n,N} $ itself.

\begin{figure}[ht]
\includegraphics[angle=0,width=1\columnwidth]{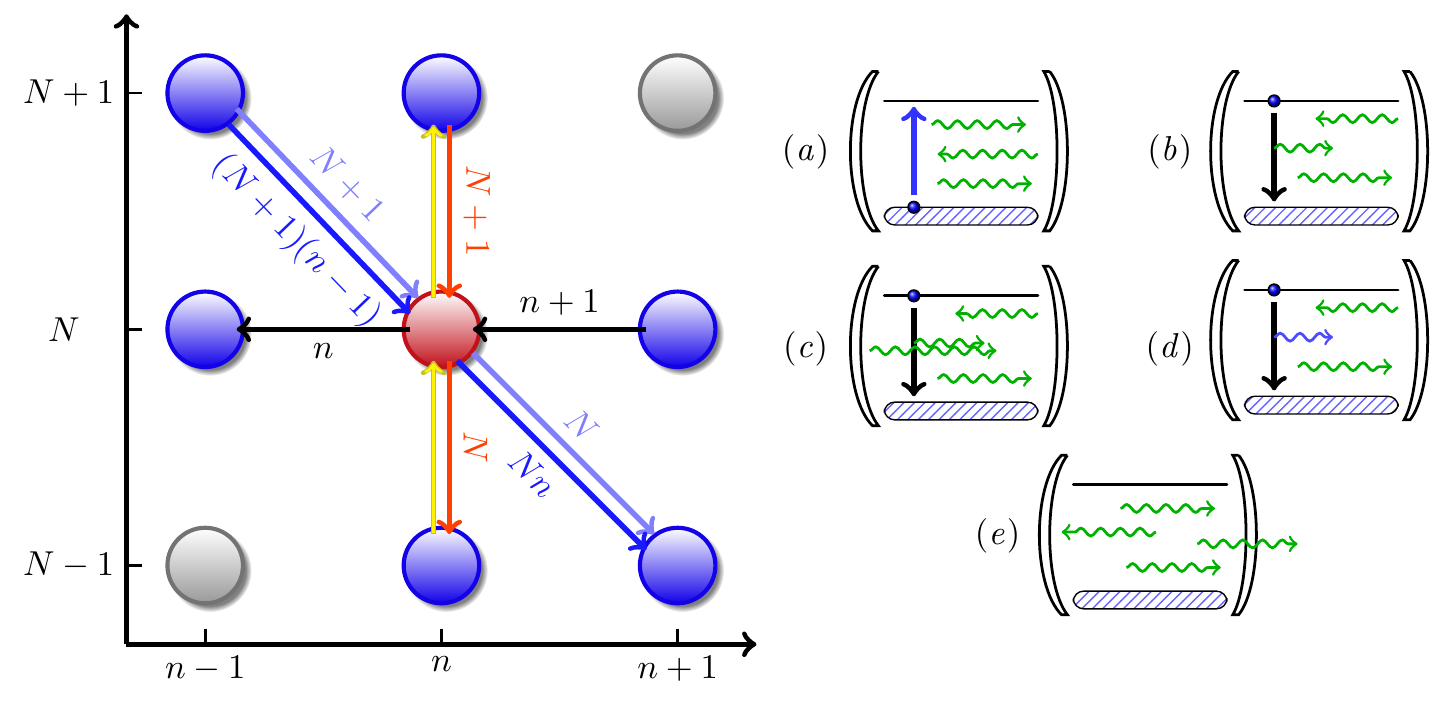}
\caption{Schematic description of the birth-death model. $n$ is  the photon number in the laser mode and $N$ the atom or carrier number in the upper level. In the left picture the processes in Eq.~\eqref{eq:mg} are symbolized for a chosen $\frac{ \operatorname{d} \ }{\operatorname{d} t} p_{n,N}$: the circles stand for states, the appearing arrows are representing  transition rates in or out (depending on arrow direction) of the state $(n,N)$, which is represented as red circle. The corresponding rate weights are written along the arrows. In the right picture the phenomenological processes listed in Table \ref{tab:phemproc} are visualized.}
\label{fig:birthdeath}
\end{figure}

In the model, five phenomenological processes, illustrated in Fig.~\ref{fig:birthdeath} and listed in Table~\ref{tab:phemproc}, are taken into account, each with its own rate normalized with the total spontaneous emission rate~$1/\tau_\text{sp}$.
\begin{table}[h]
  \centering
\begin{tabular}{lcl}\hline \hline 
Fig.~\ref{fig:birthdeath} & rate & process\\\hline
	(\emph a)& $P$ & pump \\
	(\emph b)& $\beta$ & spontaneous emission into the lasing mode\\
	(\emph c)& $\beta$ & stimulated emission into the lasing mode \\
	(\emph d)& $1-\beta$ & spontaneous emission into nonlasing modes\\
	(\emph e)& $\kappa$ & cavity losses\\	
\hline \hline 
\end{tabular}\\
\caption{Phenomenological processes in Rice and Carmichael's birth-death model.}
\label{tab:phemproc}
\end{table}
 
If one takes into account all in- and outgoing rates for each $p_{n,N}$, and especially weighs outgoing rates negatively and incoming positively, then it results the birth-death master equation~\cite{RiceCarmichael94}
\begin{align}
	\begin{aligned}
	\frac{ \operatorname{d} \ }{\operatorname{d} t} p_{n,N}=&\ -\kappa \big[ n\cdot p_{n,N}-(n+1)\cdot p_{n+1,N}\big]\\[-0.5em]
	&\ -\beta \big[nN\cdot p_{n,N}-(n-1)(N+1)\cdot p_{n-1,N+1}\big]\\
	&\ -\beta\big[N\cdot p_{n,N}-(N+1)\cdot p_{n-1,N+1}\big]\\
	&\ -(1-\beta)\big[N\cdot p_{n,N}-(N+1)\cdot p_{n,N+1}\big]\\
	&\ +P(p_{n,N-1}-p_{n,N}) \ .
\end{aligned}\label{eq:mg}
\end{align}
For the pump process $(n,N)\to (n,N+1)$, represented as up-arrows in the left part of Fig.~\ref{fig:birthdeath}, the strength of the pump process rate is $ Pp_{n,N}$. The oppositely oriented down-arrows correspond to spontaneous emission into nonlasing modes $(n,N)\to (n,N-1)$ with the strength $(1-\beta) N p_{n,N}$. Furthermore, the spontaneous and stimulated emission into the laser mode $(n,N)\to (n+1,N-1)$ with the strengths $\beta N p_{n,N}$ and $\beta nN p_{n,N}$ are represented as diagonal down-right arrows.  Finally, the cavity losses $(n,N)\to (n-1,N)$ shown as horizontal left-arrows have the strength $\kappa n p_{n,N}$.

 The spontaneous emission coupling factor $\beta$ describes the rate of spontaneous emission into the lasing mode. Correspondingly, $(1-\beta)$ describes the rate of spontaneous emission into the nonlasing modes. In the case $\beta= 1$ the mean photon number $\langle n\rangle$ increases linearly with $P$, thus no threshold as kink in the input/output curve is visible~\cite{RiceCarmichael94} [cf. Figs.~\ref{fig:comp}(a) and (b)], hence we refer to the term ''thresholdless laser''.
 
 From the master equation \eqref{eq:mg} one can obtain equations of motions (EoM) for expectation values of the mean photon number in the lasing mode $\langle n\rangle$ and mean number of excited atoms $\langle N\rangle$ by using $\langle n^\ell N^k\rangle =\sum_{n,N=0}^\infty n^\ell N^k\cdot p_{n,N}$:
\begin{align}
\begin{aligned}
\frac{\operatorname{d} \ }{\operatorname{d} t}\langle n\rangle =&\ -\kappa \langle n\rangle+\beta\langle nN\rangle+\beta\langle N\rangle \ ,\\
\frac{\operatorname{d} \ }{\operatorname{d} t}\langle N\rangle=&\ -\langle N\rangle+P-\beta \langle nN\rangle \ .
\end{aligned} \label{eq:rat1} 
\end{align} 
These equations couple via contributions representing stimulated emission $\langle n N\rangle$ to the higher-order moments, furthermore all EoMs form an infinite hierarchy of coupled equations for moments, which corresponds to the infinite number of differential equations for probabilities in Eq.~\eqref{eq:mg}. Thus, if one chooses a finite number of EoMs, then there are always more moments as variables than equations, so any such finite system is unsolvable. The most simple method to truncate the hierarchy is to use factorization approximations like setting in the first order in Eq.~\eqref{eq:rat1}: $\langle nN\rangle =\langle n\rangle \langle N\rangle$. In this way one derives the well-known laser rate equations, which was one of the motivations for introducing the birth-death model~\cite{RiceCarmichael94}. 
Of course, it is also possible to use factorization approximations on higher levels of the hierarchy. From the resulting EoM one can determine the correlation between photon and carrier number expressed by the expectation value~$\langle nN\rangle$.

\section{Implementation of the method}
\label{sec:imp}
For a successful MEM-construction one first has to know the values of chosen photon moments. While it is possible to obtain these from the above described laser rate equations or truncated EoM~\cite{LFW14}, in several optical experiments the mean photon number can be directly measured. With this given a-priori information, we first discuss the mapping from moment values to Lagrange multipliers in Sec.~\ref{sec:itnewton} and second consider some limitations on the MEM in Sec.~\ref{subsec:bounds}.   

\subsection{Iterative Newton Method}
\label{sec:itnewton}
Determining the MED basically boils down to finding the Lagrange multipliers in Eq.~\eqref{eq:mev}. A convenient way to numerically calculate the optimal Lagrange multipliers \(\hat{\lambda}\) is to solve the dual optimization problem \(\hat{\lambda} = \argmin \Gamma(\lambda)\)
as proposed in~\cite{Batou2013}. Here the objective function \(\Gamma\) reads
\begin{align}
\Gamma(\lambda) = \braket{\lambda, \mu} + \ln Z(\lambda) \ ,
\end{align}
where \(\braket{\lambda, \mu}\) denotes the conventional inner product of vectors \(\lambda = (\lambda_1, \lambda_2, \dots, \lambda_k)\) and \(\mu = (\mu_1, \mu_2, \dots, \mu_k)\). Since the Hessian matrix is positive definite, the function \(\Gamma\) is strictly convex and takes its unique minimum such that \(\nabla\Gamma(\hat{\lambda}) = 0\) holds. Starting with an initial value \(\lambda^{(0)}\) (usually \(\lambda^{(0)} = 0\) is a decent choice) we use an iterative (relaxed) Newton method with the update rule
\begin{align}
\lambda^{(i + 1)} = \lambda^{(i)} - \alpha \left[H_\Gamma(\lambda^{(i)})\right]^{-1} \cdot \nabla\Gamma(\lambda^{(i)}) \ .
\end{align}
until convergence is reached, i.e., \(\|\nabla\Gamma(\lambda^{(i)})\| \leq \varepsilon\) with a tolerance \(\varepsilon\) close to zero. The relaxation constant \(0 < \alpha < 1\) ensures convergence and the gradient \(\nabla\Gamma\) and Hessian matrix \(H_\Gamma\) are given with
\begin{align}
\nabla\Gamma(\lambda) &= \mu - \braket{x} ,\\
H_\Gamma(\lambda) &= \braket{x \otimes x} - \braket{x} \otimes \braket{x} . \label{eq:hess}
\end{align}
Here \(\otimes\) denotes the outer product and \(x\) is the vector of moments calculated with the current Lagrange multipliers \(\lambda^{(i)}\) from iteration step \(i\). Expectation values have to be evaluated component-wise, hence \(\braket{x} = (\braket{n}, \braket{n^2}, \dots, \braket{n^k})\).
In contrast to the general theory outlined in Sec.~\ref{sec:mem}, the numerical implementation always has to take place on a finite space \(\{0,1,\dots, n_\text{max}\}\).

\subsection{Bounds on Moments}
\label{subsec:bounds}
It is important that not all possible value sequences $(\mu_k)_{k=1}^\infty$ are allowed for moments. For most, there will not be {any} statistical distribution fitting these moment values. Accordingly, in so-called {moment problems}~\cite{shohat70} one investigates existence and uniqueness of the probability measure mapped from a given $(\mu_k)_{k=1}^\infty$. In our case of photon number measurement the outputs are positive, thus we have the {Stieltjes moment problem} and the corresponding necessary condition for moments. For all $n=0, 1, 2, \dots,\infty$ the following determinants of Hankel matrices must be strictly positive~\cite{shohat70}
\begin{align*}
	\left|\left(\begin{matrix}
1 & \mu_1 \cdots & \mu_{n}    \\
\mu_1 & \mu_2  \cdots & \mu_{n+1} \\
\vdots & \ddots  & \vdots &  \\
\mu_{n} &  \cdots & \mu_{2n}
\end{matrix}\right)\right| >0,\ \left|\left(\begin{matrix}
\mu_1 & \mu_2  \cdots & \mu_{n+1}    \\
\mu_2 & \mu_3  \cdots & \mu_{n+2} \\
\vdots &   \ddots & \vdots \\
\mu_{n+1}\!\!\!\!  & \cdots & \mu_{2n+1}
\end{matrix}\right)\right|>0.
\end{align*}
As a proposition, we derive in the case $n=1$ a lower bound for $\langle n^2\rangle$: $\langle n\rangle^2<\langle n^2\rangle$. Consequently, we can derive a condition for the autocorrelation function $g^{(2)}(0)$
\begin{align*}
	\langle n\rangle^2<\langle n^2\rangle =g^{(2)}(0)\langle n\rangle^2+\langle n\rangle\ \Rightarrow\ \big(1-g^{(2)}(0)\big)\langle n\rangle<1.
\end{align*}
For $g^{(2)}(0)\ge 1$ this inequality is trivially satisfied, whereas in the other case $g^{(2)}(0)< 1$ we get {an upper bound on the photon number expectation value only by knowing the autocorrelation value}
\begin{align*}
	\langle n\rangle<\frac{1}{1-g^{(2)}(0)}, \quad \text{ if }g^{(2)}(0)< 1. 
\end{align*}
This purely statistical bound may be important for the comparison of experimental results for single-photon sources.

So far, we have obtained only lower bounds on moments, however for a MED to exist it has to fulfill special upper bounds. For the continuous range $[0,\infty)$ this was shown in~\cite{Einbu77} as Theorem 2, rather one can use results in the discrete case $\mathbb N_0$, too. We formulate  these like in~\cite{Einbu77}:
{If the MED for the $k-1$ moments $\mu_1, \mu_2,\dots,\mu_{k-1}$ associated with $\lambda_1,\lambda_2,\dots,\lambda_{k-1}$, so-called $(k-1)^\text{th}$-order MED, exists, then MED for $k$ moments ($k^\text{th}$-order MED) exists only if a-priori moment $\mu_k$ is smaller than the $k^\text{th}$ moment of $k-1$-order MED:}
\begin{align}
	\mu_k\le \mu_{k,\text{\emph{max}}}:=\sum_n \frac{n^k}{Z(\lambda)} \exp\left(-\sum_{i=1}^{k-1}\lambda_i n^i\right) . \label{eq:ub} 
\end{align}
This upper moments bound, if existing, offers a criterion for gradual application of MEP. If the MEP was successful in the $(k-1)^\text{th}$-order then one can check with Eq.~\eqref{eq:ub} whether it is reasonable to try the next step $k$. Nevertheless, this $k^\text{th}$ inequality is necessary {only} for the MED of $k^\text{th}$-order. Furthermore, if the $k^\text{th}$-order is not existing, one should try MEP in the next $(k+1)^\text{th}$-order, where MED might still exist.

For the second-order MED $\exp(-\lambda_1n-\lambda_2n^2)/Z(\lambda)$ to exist, it results from Eq.~\eqref{eq:ub} that the Dowson-Wragg inequality~\cite{dowwrag70} $\langle n^2\rangle\le 2\langle n\rangle^2$ is required, because 
the first-order MED always exists for $\mu_1>0$, since for $\exp(-\lambda_1n)/Z(\lambda)$ follows $\lambda_1=\ln(1+1/\mu_1)>0$.
Furthermore, for the first-order MED the following applies: $\langle n^2\rangle= 2\langle n\rangle^2$.
 Consequently, like mentioned in~\cite{lettau2018superthermal} only in the case $g^{(2)}(0)\le2$, it may be possible to find a second-order MED, but it is also possible to find a third-order MED with $g^{(2)}(0)>2$. 

 In numerical implementations of this method, we always choose an approximation space $\{0,1,\dots, n_\text{max}\}$  instead of the configuration space $\mathbb N_0$ to work with. In this finite situation there are no upper bounds like Eq.~\eqref{eq:ub} and especially there are second-order MEDs with $g^{(2)}(0)>2$ for $n_\text{max}$ big enough~\cite{Tag00}. The necessary and sufficient conditions for the existence of a maximum entropy solution are identical to the general ones for the finite moment problem. This means, that if one measures or calculates moments for a finite range, which for numerical applications is the usual case, then the MED exists for this {finite range}, even though the MED for the corresponding infinite range may not exist. 

\section{Numerical Results}
\label{sec:num}
We first test in Sec.~\ref{subsec:photrec} the MEM construction of stationary photon distributions  and discuss some appearing issues. Then in Sec.~\ref{subsec:defth} we introduce a new characterization of the emitted light using the entropy and the first Lagrange multiplier. After that, in Sec.~\ref{subsec:entinf} we compare the entropy curve to the entropy of other distributions and additionally handle the consequences of $g^{(2)}(0)\approx 1$. Finally, in Sec.~\ref{subsec:fullrec} the MEM is applied to determine the full statistics of photons and carriers.

\subsection{Photon Distribution Construction} 
\label{subsec:photrec}
In contrast to our original motivation, we here do not use moments from EoM but instead we determine the moment values $\mu_k=\langle n^k\rangle$ directly from the steady-state photon distribution $p_n$ of the birth-death model. By doing so, the performance check of the MEM is not mixed up with truncation errors of a given EoM. The latter are already discussed in detail in the literature (see, e.g.,~\cite{LFW14}). In the following, we present results based on numerically constructed MEDs up to the tenth order. 

\begin{figure}[h]
	\includegraphics[width=1\columnwidth]{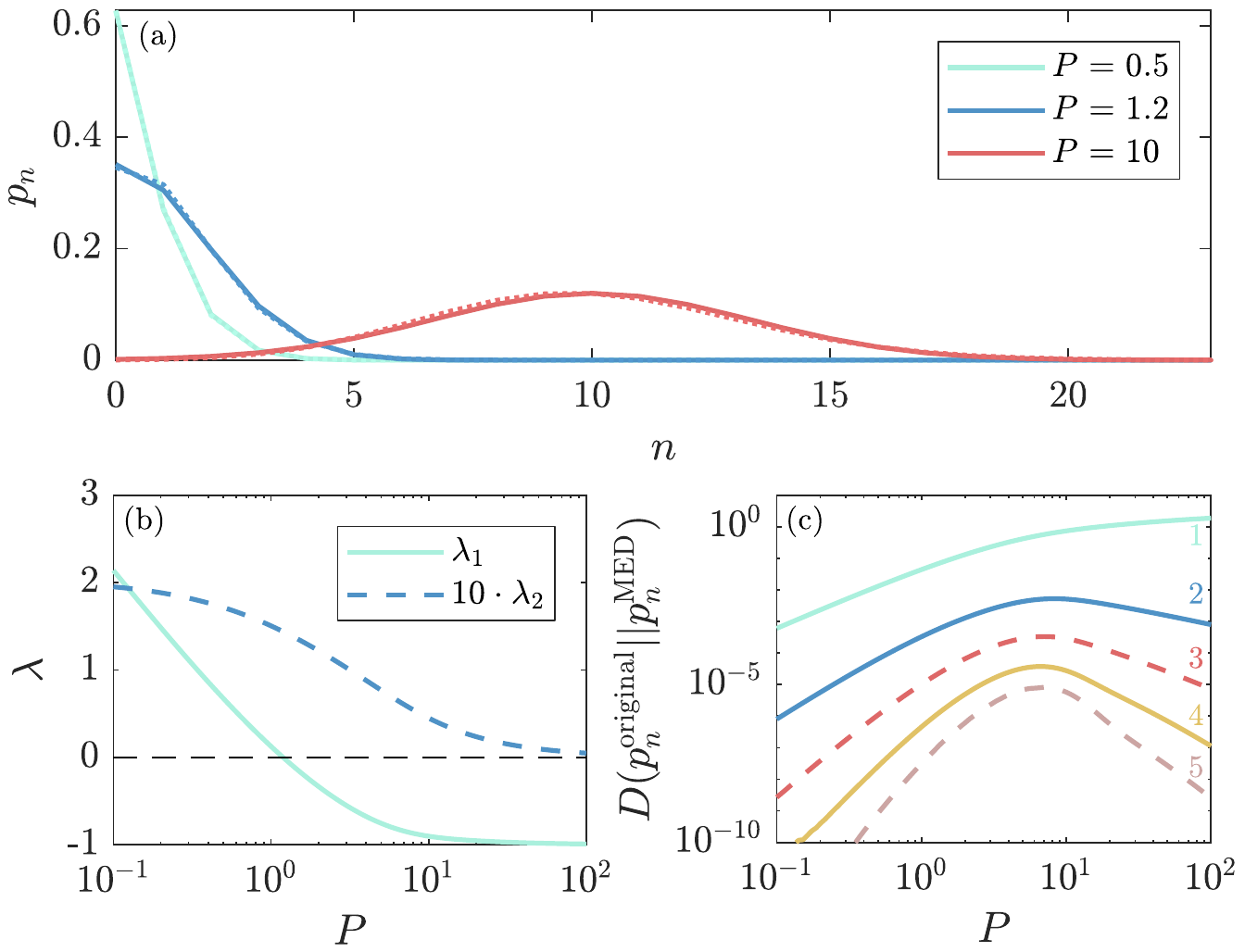}
	\caption{(a) Photon statistics $p_n$ for pump rates~$P$ (like all rates normalized by the total spontaneous emission rate~$1/\tau_\text{sp}$) well below, at and well above the laser threshold as characterized by the sign change in the first Lagrange multiplier \(\lambda_1\) [see (b) and Sec.~\ref{subsec:defth}]. Solid curves show the MED of second order, dashed curves the original distributions. Although the distributions nearly lie on top of each other, higher orders of MEM lead to even better agreement. (c) Solid curves show the Kullback divergence in Eq.~(\ref{eq:kullback}) for MEDs that exist on \(\mathbb{N}_0\), dashed curves indicate MEDs (here third and fifth order) that do only exist for finite photon numbers. The parameters are $\kappa = 1 = \beta$.}
	\label{fig:KLD}
\end{figure}

Figure~\ref{fig:KLD}(a) shows the comparison of the original birth-death model distributions and the second-order MEDs. Surprisingly, for all three values of the pump rate the constructed distributions are almost identical to the original curves. So the photon distribution has nearly the Gaussian form $\exp(-\lambda_0-\lambda_1 n- \lambda_2 n^2)$ in this regime of pump rates. Due to the bounds discussion in Sec.~\ref{subsec:bounds}, this indicates that  $g^{(2)}(0)\le2$ is also true for the original photon distribution.

The approximation quality, measured by the Kullback divergence~\cite{Cover1991}
\begin{align}
	D(p^\text{original}\|p^\text{MED})=\sum_i p^\text{original}_ i \ln\left(\frac{p^\text{original}_i}{p^\text{MED}_i}\right), \label{eq:kullback}
\end{align}
is getting better with higher order [Fig.~\ref{fig:KLD}(c)], hence the sequence of MEDs converges to the original distribution. We use $D(p^\text{original}\|p^\text{MED})$ because it measures the informational inefficiency of choosing  $p^\text{MED}$ instead of the original distributions $p^\text{original}$. Moreover, with the Kullback divergence we are able to estimate the summed absolute distance as $2D(p||q)\ge \sum_i|p_i-q_i|$~\cite[p.300]{Cover1991}. In Fig.~\ref{fig:KLD}(c) we can observe that for each order the highest errors lie in the pump rate transition range between the lasing and the nonlasing regime,
as in this range the original distribution is the most complicated one.

The behavior of the calculated MED depends strongly on whether it is an even or odd order. The even-order MEDs exist also on the global range $\mathbb N_0$, because each last Lagrange multiplier $\lambda_\text{order}$ is positive, so that from a certain point on the MEDs are rapidly decreasing for higher photon numbers. In addition, all Lagrange multipliers do not depend on the chosen size of the approximation space $n_\text{max}$. In this sense the even-order MEDs are well defined approximations.

\begin{figure}[h]
	\includegraphics[width=1\columnwidth]{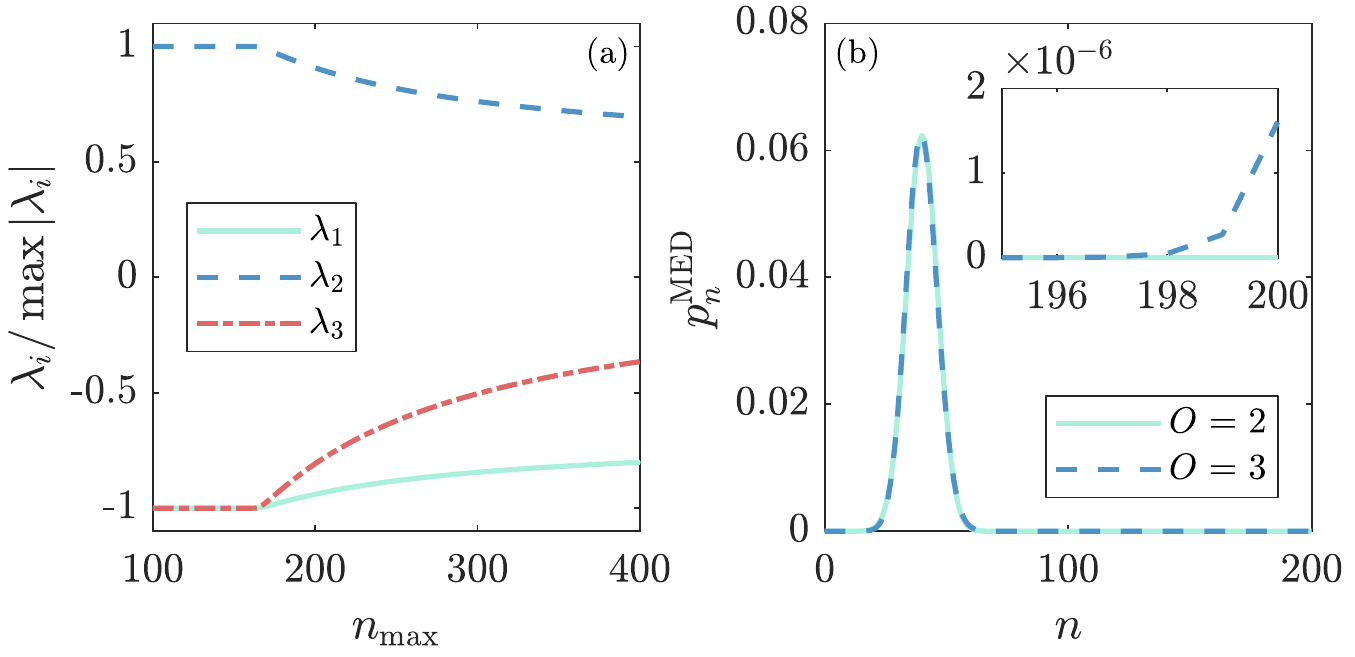}
	\caption{(a) Normalized Lagrange multipliers for MEM of third order for different maximum photon numbers~\(n_\text{max}\). Even orders of MEM lead to \(n_\text{max}\)-independent Lagrange multipliers (not shown). In contrast, odd orders produce \(n_\text{max}\)-dependent Lagrange multipliers which are constant for small \(n_\text{max}\) and tend to the values of the previous order for \(n_\text{max}\) approaching infinity. Furthermore, in the odd-order case, the MED (b) exhibits increasing values for photon numbers close to the maximum value \(n_\text{max}\) (inset). The parameters are in (a)-(b) \(P = 40\), \(\kappa = 1 = \beta\) and in (b) \(n_\text{max} = 200\).}
	\label{fig:lamb_vs_nmax}
\end{figure}

For the odd MEM-orders each $\lambda_\text{order}$ has a negative sign and all Lagrange multipliers depend strongly on  $n_\text{max}$. By the discussion in the end of Sec.~\ref{sec:mem}, it becomes clear, that the odd MEDs cannot be expanded from the range   $\{0,1,\dots, n_\text{max}\}$ to the true configuration space $\mathbb N_0$. In contrast to the even orders, for the odd orders the MED is growing at the border $n_\text{max}$ to a slope (inset of the right panel of Fig.~\ref{fig:lamb_vs_nmax}). All this implies that the MED is not existing on $\mathbb N_0$ for odd orders. 
In fact, we can also conclude this non-existence by testing the highest moment with the inequality \eqref{eq:ub}, because for the odd orders this necessary condition is violated, see Table~\ref{tab:ineqtest}.

\begin{table}[h]
\begin{tabular}{cccc}\hline \hline 
order & $\mu_\text{{order}}\cdot 10^{-\text{{order}}}$ & $\mu_\text{{order,max}}\cdot 10^{-\text{{order}}}$   & existence on $\mathbb N_0$\\\hline
	1& 1.00000000 & $+\infty$& yes\\
	  2& 1.10996480 & 2.10000000 & yes\\ 
	   3&  1.33934846
	    & 1.33080223& no \\ 
	    4&   1.73473327 & - &yes \\
   5& 2.39038816 & 2.39011279 & no \\ 
   6& 3.48107450 & - & yes \\
   7& 5.32975586 & 5.32971709 & no \\
   8 &8.54326181 & - & yes \\
  9& 1.42872866 & 1.42872768 & no  \\ 
  10& 2.48546984 &- & yes \\\hline\hline
\end{tabular}
\caption{Moments $\mu_k$ listed by order in comparison to maximal possible moment values $\mu_\text{{$k$,max}}$ from inequality~\eqref{eq:ub}  for MED existence on a global range $\mathbb N_0$  for $P=10$, $\kappa = 1 = \beta$ . For some orders $\mu_\text{{$k$,max}}$ is lacking, because due to inequality \eqref{eq:ub} the MED of the previous order on $\mathbb N_0$ does not exist.}
\label{tab:ineqtest}  
\end{table}

In Table~\ref{tab:ineqtest} the value of the second moment $\mu_2$ is smaller than the second moment of the first-order MED $\mu_{2,\text{max}}$ that corresponds to the existence of the second-order MED. Since the odd-order moments are strictly greater than allowed by $\mu_\text{{order,max}}$, the odd-order MEDs cannot exist on $\mathbb N_0$. This circumstance explains why the  odd-order MEDs on approximation  spaces $\{0,1,\dots,n_\text{max}\}$ do not converge for $n_\text{max}\to \infty$ to a MED on $\mathbb N_0$. Also in this way the $n_\text{max}$-dependence of Lagrange multipliers, the slope on the space-limit $n_\text{max}$ and the negative sign of the last $\lambda_\text{order}$ is clarified. 

Intriguingly, for each odd order $O=2k+1$, $k\in \mathbb N_0$ in the limit $n_\text{max}\to \infty$ the $O^\text{th}$ order MED converges to the MED of the previous even  $(O-1)^\text{th}$ order, which exists globally. Thus, $\lambda_O \nearrow 0 $ and other Lagrange multipliers converge to previous-order Lagrange multipliers. The arising issue is that the $O^\text{th}$ moment value $\mu_\text{order}$ of odd $O^\text{th}$ order MEDs is strictly greater than the $O^\text{th}$ moment value $\mu_\text{order, max}$ of the $(O-1)^\text{th}$-order MED (Table~\ref{tab:ineqtest}). Effectively, for $n_\text{max}\to \infty$ this error $\mu_\text{order}-\mu_\text{order, max}$ is compensated by the distribution slope at $n_\text{max}$. At the same time the probability weight of the slope falls $\propto n_\text{max}^{-O}$, so that ($O=2k+1$)
\begin{align*}
	\sum_{i=1}^{n_\text{max}}\left|p^\text{$O^\text{th}$ MED}_i-p^\text{$(O-1)^\text{th}$ MED}_i\right| \xrightarrow {n_\text{max}\to\infty} 0 \ .
\end{align*}

Furthermore, for the odd orders one in general observes a kink in $ D(p^\text{original}\|p^\text{MED})$ (not shown) and in all Lagrange multipliers $(\lambda_i)_i$ at the same point $n_\text{kink}$ [Fig.~\ref {fig:lamb_vs_nmax}(a)]. For the lower $n_\text{max}$ up to $n_\text{kink}$, all $\lambda_i=\Lambda_i\in \R$ stay constant, and from $n_\text{kink}$ on these quantities behave for $n_\text{max}\to \infty$ in the way described above. 

The constant values up to the  kink can be explained as follows. In the example visualized in Fig.~\ref{fig:lamb_vs_nmax}(b) the third-order MED can be generally separated into two parts: the Poisson-like peak and the slope at the limit of the range. So it may happen that the Poisson-like peak already fulfills the moment constraints comparable to case $n_\text{max}=160$. Nevertheless, the range limit is far away from the peak, and additionally the slope at $n_\text{max}$ is so small that it does not influence the moment values. Let 
\begin{align}
	\exp(-\Lambda_1 n -\Lambda_2 n^2-\Lambda_3 n^3)/Z(\Lambda) \label{eq:exp3}
\end{align}
be the MED for $n_\text{max}=160$ with Lagrange multipliers $(\Lambda_1,\Lambda_2,\Lambda_3)$.
 Now, if one restricts the MED \eqref{eq:exp3} as mathematical function  to $\{0,1,\dots,n_\text{max}=140\}$ by leaving the Lagrange multipliers invariant ($\lambda_i=\Lambda_i\in \R$ for $i=1,2,3$), then the moment constraints are still fulfilled. The reason is that in the remaining range $\{141,\dots,160\}$ the probabilities  are almost zero, thus they do not contribute either to the moment values or the partition function $Z(\lambda)$. Hence, the MED for $n_\text{max}=140$ is the MED for $n_\text{max}=160$ truncated at $n=140$. As a consequence, the Lagrange multipliers stay constant below $n_\text{max}=160$.

\begin{figure}[h]
	\includegraphics[width=1\columnwidth]{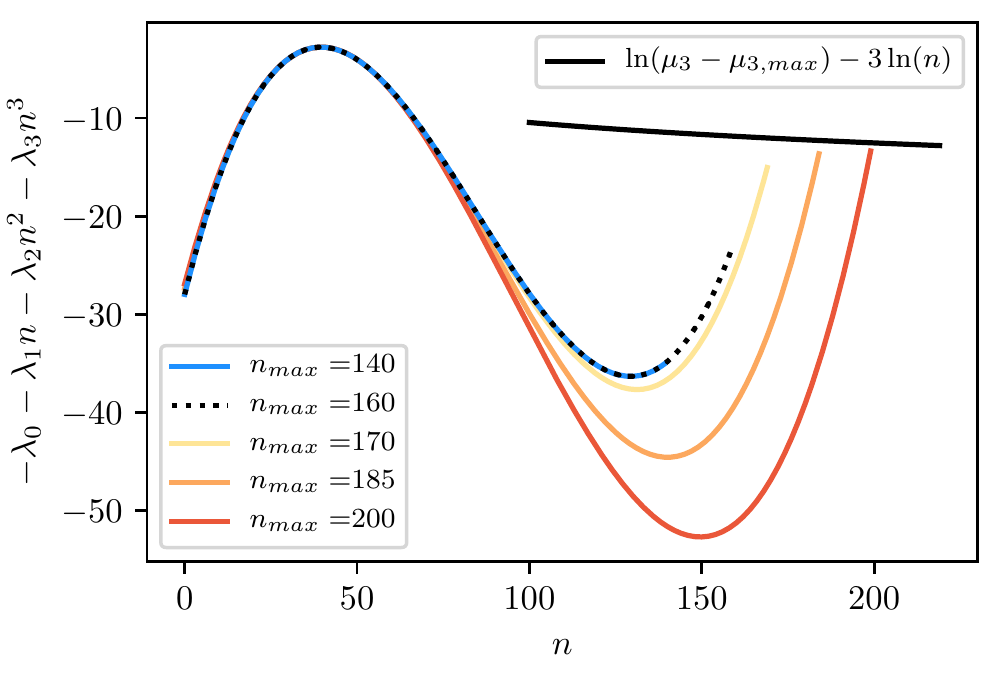}
	\caption{Exponent of the third-order MED at $P=40$, $\kappa=1 = \beta$ for different approximation space ranges $n_\text{max}$. Up to $n_\text{max}=160$ the MEDs are restrictions of the same MED with the fixed Lagrange multipliers $(\Lambda_1,\Lambda_2,\Lambda_3)$. The kink $n_\text{kink}$ [see Fig.~\ref{fig:lamb_vs_nmax}(a)] appears at the point where the MED with $(\Lambda_1,\Lambda_2,\Lambda_3)$ would cut the curve $\ln (\mu_{3}-\mu_\text{3,max})-3 \ln(n) $ obtained from Eq.~\eqref{eq:kink}.}
	\label{fig:exp_kink}
\end{figure}

With higher $n_\text{max}$ the kink appears because for the MED with $(\Lambda_1,\Lambda_2,\Lambda_3)$ the slope would explode for $n>200$. To shift this blow up beyond of the range $\{0,1,\dots,n_\text{max}\}$, $|\Lambda_3|$ gets smaller.  Moreover, for $n_\text{max}\to\infty$   the slope at $n_\text{max}$ is getting tighter, since $p_{n_\text{max}}$ gets much greater then the previous probability $p_{n_\text{max}-1}$ in the same MED. So, for greater $n_\text{max}$ the slope is fully determined by $p_{n_\text{max}} $. Furthermore, it should be noted that $p_{n_\text{max}}$ gets large enough so it can correct the moment error $\mu_\text{order}-\mu_\text{order, max}$, e.g.,
\begin{align}
	\mu_\text{3}-\mu_\text{3, max} \approx n_\text{max}^\text{3}\cdot p_{n_\text{max}} \Rightarrow p_{n_\text{max}}
	\approx \frac{ \mu_{3}-\mu_\text{3,max}}{(n_\text{max})^3} \label{eq:kink} \ .
\end{align}
From the above formula \eqref{eq:kink} it also results $n_\text{max}^\text{1}\cdot p_{n_\text{max}}\to 0$ and $n_\text{max}^\text{2}\cdot p_{n_\text{max}}\to 0$, so that the slope does not have much influence on the first and second moments.
At the same time, the Poisson-like part of the third-order MED converges to the second-order MED. 
Thus, as shown in Fig.~\ref{fig:exp_kink}, the kink position $n_\text{kink}$ is determined by the crossing point between the $-\Lambda_0-\Lambda_1 n -\Lambda_2 n^2 - \Lambda_3 n^3$ and the curve $\ln (\mu_{3}-\mu_\text{3,max})-3 \ln(n_\text{max}) $ from Eq.~\eqref{eq:kink}, which describes the behavior of $p_{n_\text{max}}$ for high ${n_\text{max}}$.

In conclusion, if one only takes the Poisson-like part of a third-order MED, thus restricts it on $\{0,1,\dots,n_\text{kink}\}$, then it is possible to consider this third-order MED as well defined. Namely, it is valid that first below $n_\text{kink}$ the deviation $ D(p^\text{original}\|p^\text{MED})$  is smaller than above the kink, and second the Lagrange multipliers stay constant up to the kink. Similarly, the analogous results can be derived 
for the higher odd orders
\begin{equation}
p_{n_\text{max}}
	\approx \frac{ \mu_\text{order}-\mu_\text{order,max}}{(n_\text{max})^\text{order}} \ .
\end{equation}



\subsection{Characterization of the emitted light} 
\label{subsec:defth}
In this subsection we introduce a new characterization of the emitted light by the entropy and the first Lagrange multiplier. We demonstrate that the latter can be used to distinguish between the nonlasing and the lasing regime. This is usually done by observing the kink in the intensity, the steep increase in the coherence time, and the step in the autocorrelation function $g^{(2)}(0)$ as function of the pump rate~\cite{Ulrich06,Wiersig10}. Using these quantities a clear laser threshold can only be located for small values of the spontaneous emission coupling factor $\beta$~\cite{RiceCarmichael94}. For $\beta$ close to unity the kink in the intensity disappears and the step in the autocorrelation function is smeared out. Also, higher-order autocorrelation functions indicate that it might be better to speak about a threshold region rather than a threshold point~\cite{LFW14,FLW17}. However, we show that the first Lagrange multiplier allows to define, at least on a formal level, an unambiguous threshold point. 

\begin{figure}[h]
\includegraphics[width=1\columnwidth]{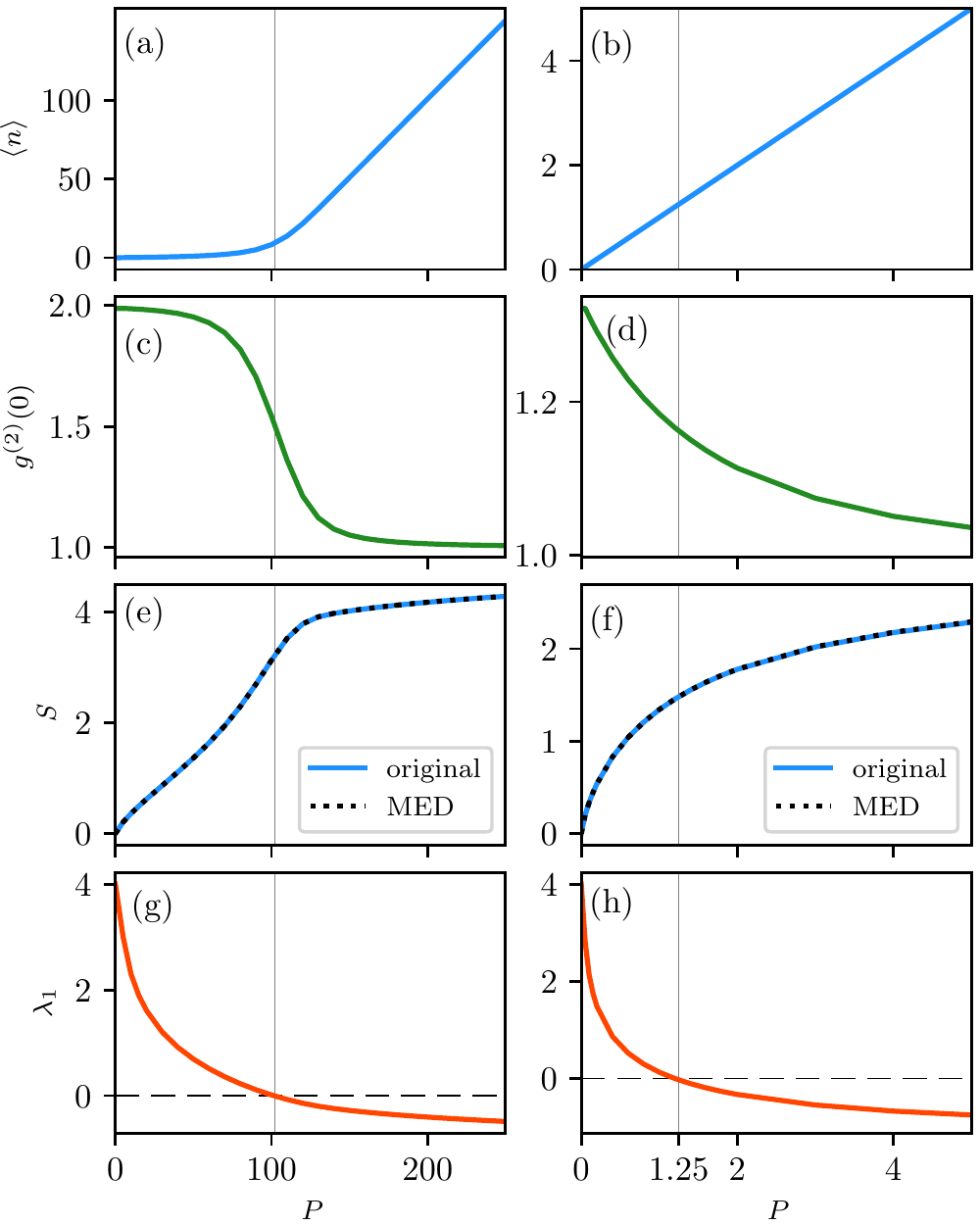}
\caption{Comparison of possible threshold characterizations for $\kappa=1$ with $\beta=0.01$ in (a), (c), (e), (g) and $\beta=1$ in (b), (d), (f), (h) by the photon number in (a)-(b), the autocorrelation function $g^{(2)}(0)$ in (c)-(d), the original photon entropy and the photon entropy of MEM-approximation $S$ in (e)-(f), and the first Lagrange multiplier $\lambda_1$ in (g)-(h). Vertical lines mark the pump rate~$P$ where $\lambda_1$ is zero.}
\label{fig:comp}
\end{figure}

For the numerical calculation, we choose two values of the spontaneous emission coupling factor, $\beta=0.01$ with a kink in the input-output curve [Fig.~\ref{fig:comp}(a)], and $\beta=1$, where the photon number (proportional to the detected intensity) increases linearly with the pump power [Fig.~\ref{fig:comp}(b)]. In the latter case no threshold can be identified in the photon number. 
For $\beta=0.01$, the autocorrelation function jumps in Fig.~\ref{fig:comp}(c) from $g^{(2)}(0)\approx 2$, indicating thermal light, to $g^{(2)}(0)\approx 1$, indicating coherent light. For $\beta=1$ the jump is smeared out in Fig.~\ref{fig:comp}(d), so the threshold still cannot be identified.

The entropy~$S$ increases with the pump rate~$P$ in both cases, see Fig.~\ref{fig:comp}(e) and (f), indicating that the photon distribution is getting broader. Analogous to the observations in~\cite{BP89} for a single-mode Scully-Lamb theory, there is a kink in the entropy for $\beta=0.01$ in Fig.~\ref{fig:comp}(e), but not for $\beta=1$ in Fig.~\ref{fig:comp}(f). 
The better way is to extract the photon distribution properties  directly from the first two moments via the MEM. The transition from a thermal distribution below the threshold to a Poisson distribution in the lasing regime can be described by the transformation from a monotonically decreasing function to a peaked one~\cite{SL67}. At some $P_\text{th}$ the photon distribution develops an extremum, when $p_n$ has a zero slope in $n=0$. If one chooses as approximation the second-order MED (Gaussian) then the existence of the extremum is equivalent to
\begin{align*}
	&\frac{\operatorname{d}\ }{\operatorname{d} n} \exp\left(-\lambda_{1}\cdot n-\lambda_{2}\cdot n^{2}\right) = 0 \\
	& \Rightarrow \quad \lambda_{1}+2\lambda_{2}\cdot n =0 \quad \Rightarrow \quad \lambda_1\le 0,
\end{align*}
because $n$ and $\lambda_2$ are both nonnegative. As a result, at the pump power $P_\text{th}$ with $\lambda_1=0$ the MED develops a zero slope at the first time. Especially, above $P_\text{th}$ the multiplier becomes $\lambda_1<0$, so the photon distribution is a peaked Gaussian. 

Consequently, we define the threshold pump power~$P_\text{th}$ by the condition $\lambda_1=0$. For $\beta=0.01$ it is shown in Fig.~\ref{fig:comp}(a), (c), (e), and (g) that $P_\text{th}$ (visualized as vertical line) is consistent to the other threshold definitions by the photon number in (a), $g^{(2)}(0)$ in (c), and the entropy in (e). Moreover, in case of $\beta=1$ this new $P_\text{th}$ definition by $\lambda_1$ is the only possible one, because photon number, $g^{(2)}(0)$, or entropy do not possess any kinks or jumps. 
As already mentioned above, for large~$\beta$ the concept of a laser threshold has been criticized~\cite{RiceCarmichael94}. However, we believe that our definition of a threshold by $\lambda_1 = 0$ is useful as this condition marks a qualitative change in the photon statistics.

\begin{figure}[h]
\includegraphics[width=1\columnwidth]{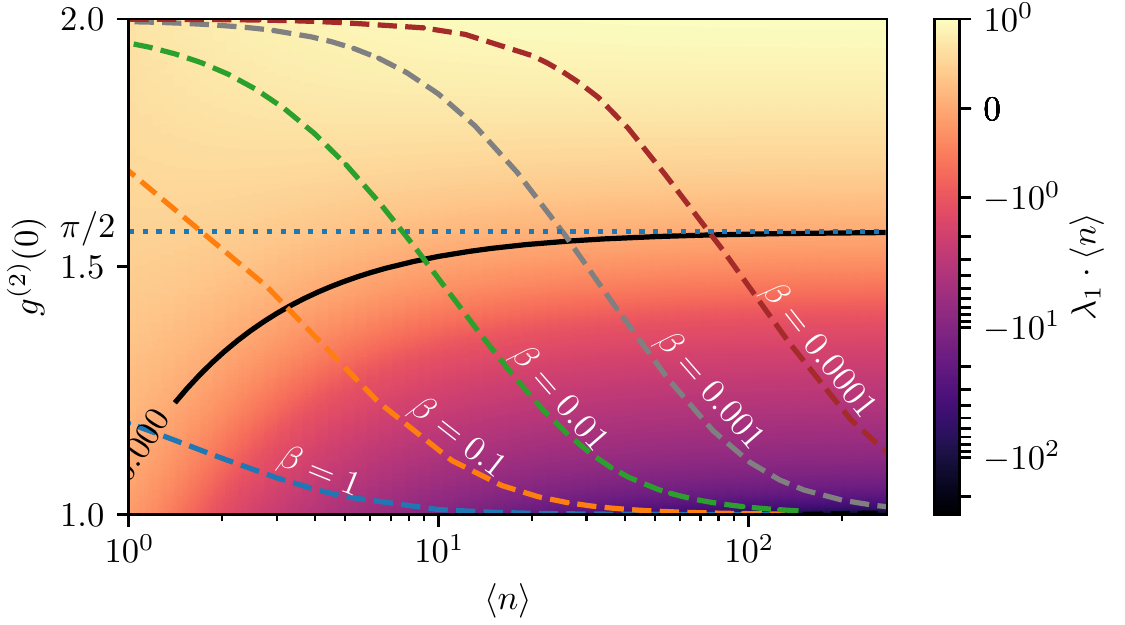}
\caption{The first Lagrange multiplier $\lambda_1$ of the second-order MED with given $\langle n \rangle$ and $g^{(2)}(0)$ values. The zero level of $\lambda_1$, where the photon statistics changes its character, is drawn as the solid curve. It converges monotonically to $\pi/2$, represented by the dotted line, as the photon expectation value $\langle n\rangle$ increases. Additionally, the dependency between $\langle n\rangle$ and $g^{(2)}(0)$ of the birth-death model for $\beta=1, 0.1, 0.01, 0.001$ and $0.0001$ is shown as dashed curves.}
\label{fig:thresholdlambda}
\end{figure}

To show the usability of this threshold definition, we relate the condition $\lambda_1=0$ to a condition on $g^{(2)}(0)$, which is directly observable. In fact, the determination of the $\lambda_1$ value also requires the knowledge of the first moment, the photon number $\langle n\rangle$, whereas in experiments only the detected intensity is known. The value of $\langle n^2\rangle$ is also needed for the second-order MED, but can be directly calculated out of the given $\langle n\rangle$ and $g^{(2)}(0)$ values. To avoid the necessity of $\langle n\rangle$, which is usually lacking, we calculate for each value of $\langle n\rangle$ the $g^{(2)}(0)$ value where $\lambda_1=0$. In Fig.~\ref{fig:thresholdlambda} this $(\lambda_1=0)$-curve separates the region with $\lambda_1>0$ corresponding to a thermal distribution and $\lambda_1<0$ corresponding to a peaked distribution. Also the $(\lambda_1=0)$-curve converges for $\langle n\rangle \to \infty$ to the value $\pi/2\approx 1.571$, which can be easily calculated via continuous integrals over the $\exp(-\lambda_2 n^2)$ distribution. Moreover, for the case $\langle n \rangle > 30$ the $\lambda_1=0$-curve is almost identical to $g^{(2)}(0)=\pi/2$. Thus, if we know that the phase transition will happen at higher $\langle n \rangle > 30$ then $g^{(2)}(0)=\pi/2$ will be a good criterion for the threshold. For the birth-death model it is the case for $\beta\le 0.001$, where the relation $g^{(2)}(0)=\pi/2$ provides more precise definition than the smooth jump from $g^{(2)}(0)=2$ to $g^{(2)}(0)=1$. Yet, by our definition in the more general case $\langle n\rangle\ge 1$ the least possible value of the phase transition is $g^{(2)}(0)=1.1$.


\subsection{Distribution Comparison by Entropy}
\label{subsec:entinf}
One can recover more information from the form of the photon entropy curve by comparison with characteristic Poisson, Gaussian and thermal entropy values. Below the threshold power $P_\text{th}$ one expects that a thermal distribution $p^\text{therm}_n=\exp(-\lambda_1 n)/Z(\lambda_1)$ fits the photon distribution well. If one chooses it with the same $\langle n\rangle$ as that of the photon distribution, then $(p^\text{therm}_n)_{n=0}^\infty$ is the first-order MED. It is straightforward to derive its entropy value  determined only by $\langle n\rangle$, cf. Table~\ref{tab:entrdist}.
In the lasing regime, one expects the Poisson distribution $p^\text{Poisson}_n={\langle n\rangle^n\exp(-\langle n\rangle)}/{n!}$ with the same expected photon number as a good approximation. Its entropy for large $\langle n\rangle$, listed in Table~\ref{tab:entrdist},   depends  only on $\langle n\rangle$.

Moreover, it is worth to add the Gaussian  $p^\text{Gaussian}_n={\exp(-\lambda_1 n-\lambda_2 n^2)}/{Z(\lambda)}$ to the  comparison, because the deviations of the MED in the second order from the photon distribution are small and further for  large values of $\langle n\rangle$, the Poisson distribution gets more similar to a Gaussian with expectation value $\langle n\rangle$ and variance $\text{var}(n)=\langle n^2\rangle-\langle n\rangle ^2\overset{\text{\tiny{!}}}=\langle n\rangle$. The entropy for the continuous normal distribution on the range $(-\infty,\infty)$ depends only on its variance (Table~\ref{tab:entrdist}), hence, we use this simple form to plot the Gaussian entropy directly from the photon distribution variance.
\begin{table}[h]
\begin{tabular}{ccc}\hline \hline
{distribution} & $p_n$& {entropy}\\[0.1em]\hline
	thermal & $\frac{\exp(-\lambda_1 n)}{Z(\lambda_1)}$ & 
	$ -  \ln\langle n\rangle^{\langle n\rangle}+ \ln\langle n +1\rangle^{\langle n+1\rangle}$ \\
	Poisson & $\frac{\langle n\rangle^n\exp(-\langle n\rangle)}{n!}$ & $\approx\frac{1}{2}\ln(2\pi e \langle n\rangle)$ \\
	Gaussian & $\frac{\exp(-\lambda_1 n-\lambda_2 n^2)}{Z(\lambda)}$ & $\approx\frac{1}{2}\ln[2\pi e \text{var}(n)]$ \\[0.4em]
	\hline \hline
	\end{tabular}
\caption{Distributions and their entropies. For Poisson distribution the entropy approximation for large $\langle n\rangle$ is given with $e = \exp(1)$. The Gaussian entropy is approximated by the continuous normal distribution entropy on $(-\infty, \infty)$.}
\label{tab:entrdist}
\end{table}

\begin{figure}[h]
\includegraphics[width=1\columnwidth]{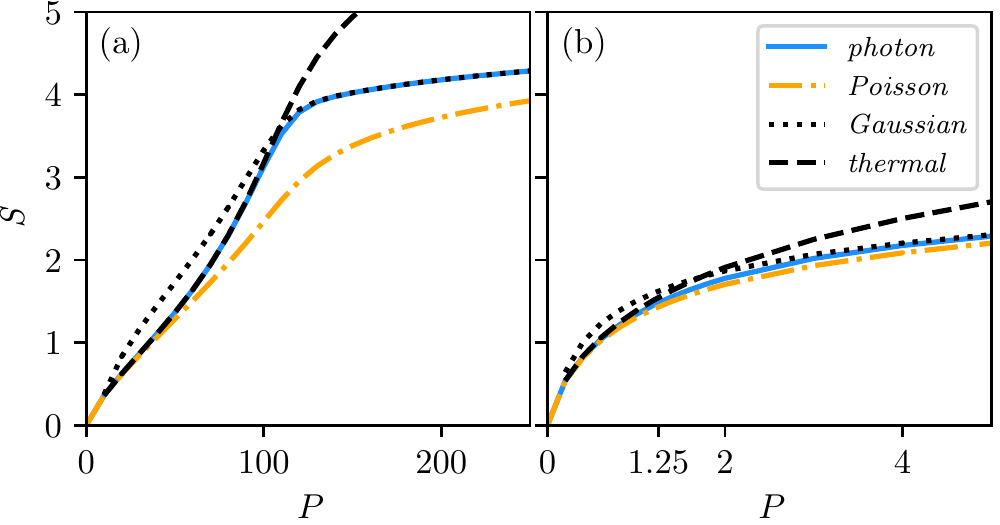}
\caption{The entropy $S$ of the photon distribution vs pump rate~$P$ for $\kappa=1$ in comparison to exact Poisson and thermal distribution entropy values with the same expectation value, as well as approximated Gaussian values with the same variance. (a) $\beta = 0.01$ and (b) $\beta = 1$.}
\label{fig:ent}
\end{figure}

In the case $\beta=0.01$ [Fig.~\ref{fig:ent}(a)] for the pump powers below $P_\text{th}$ the thermal entropy fits the photon entropy very well, however, above the threshold the Gaussian approximation is best. Additionally, in the case $\beta=1$, depicted in Fig.~\ref{fig:ent}(b), the Poisson entropy is also a good approximation of the photon entropy for $P> P_\text{th}$. If one compares these plots with Fig.~\ref{fig:comp}, then one is likely to find that in both cases $P_\text{th}$ is close to the crossing point between each Gaussian and thermal approximation in Fig.~\ref{fig:ent}. Thus, it is possible to indicate the position of $P_\text{th}$ by such a crossing. 

\begin{figure}[h] 
\vspace{1em}
\includegraphics[width=1\columnwidth]{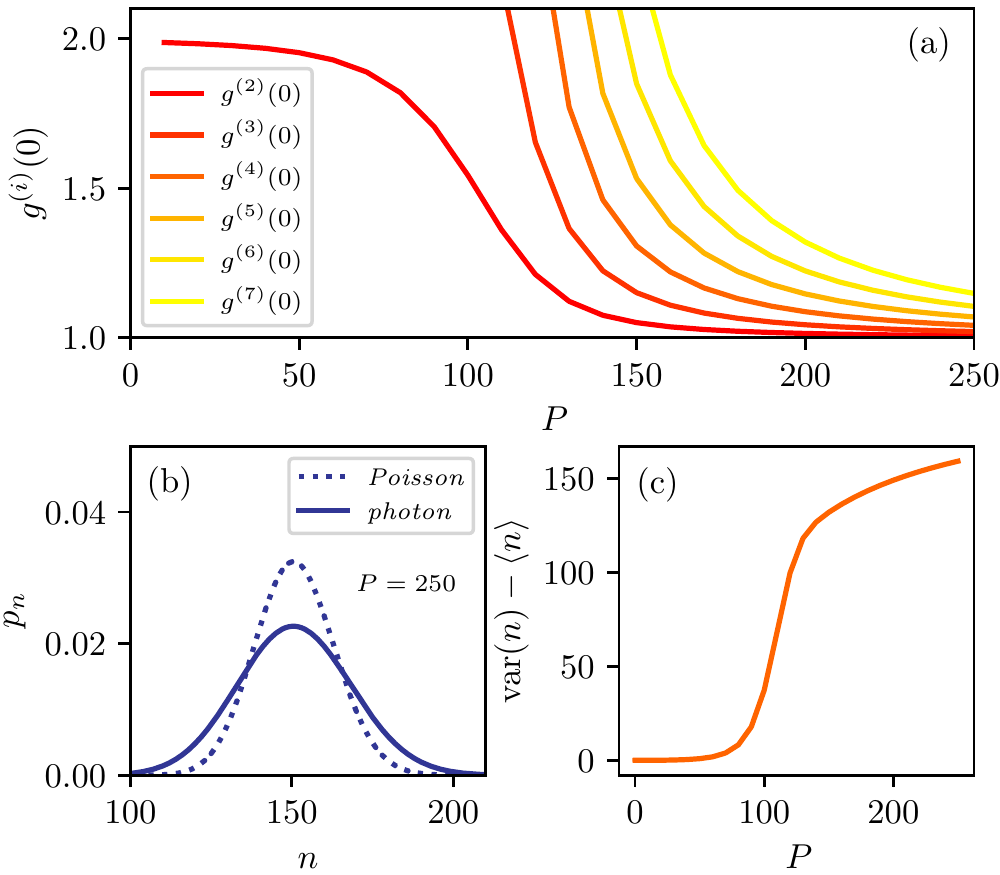}
\vspace{-2em}
\caption{(a) The autocorrelation functions $g^{(i)}(0)$ of various orders $i$ for $\kappa=1$ and $\beta=0.01$ vs pump rate $P$. (b) The broader photon and Poisson distribution $p_n$ with the same expected photon number $\langle n\rangle$. (c) The difference between variance $\text{var}(n)$ and $\langle n\rangle$.}
\label{fig:g2disc}
\end{figure}

Surprisingly, Fig.~\ref{fig:ent}(a) shows that for pump powers above threshold the Poisson entropy deviates from the photon entropy. This is unexpected because $g^{(2)}(0)\approx 1$ in the lasing regime indicates that the photon distribution should be a Poisson distribution. 
Despite the fact that the first seven autocorrelation functions $g^{(n)}(0)$ for $\beta=0.01$ in Fig.~\ref{fig:g2disc}(a) are almost equal to unity at $P=250$, the photon distribution is still much broader than the Poisson distribution with the same expectation value [Fig.~\ref{fig:g2disc}(b)], in accordance with the Poisson entropy discrepancy to the photon entropy in Fig.~\ref{fig:ent}(a). This circumstance  can be explained as follows. $g^{(2)}(0)\approx 1$ does not mean that
 \begin{align}
	\text{var}(n)-\langle n\rangle =\langle n\rangle ^2 (g^{(2)}(0)-1)\approx 0.
\end{align} 
An equality in the upper line would be one property of the Poisson distribution. In fact, for our system $\text{var}(n)-\langle n\rangle$ grows with increasing $P$ [Fig.~\ref{fig:g2disc}(c)], because $\langle n\rangle ^2 $ grows faster than $g^{(2)}(0)$ converges to 1. Thus, since $\langle n\rangle^k$  appears in $k$-autocorrelation function denominators, the Poisson characterization of probability distributions by $g^{(k)}(0)\approx 1$ is subtle, because each equation $g^{(k)}(0)= 1$ and $g^{(k)}(0)\cdot \langle n\rangle^k= \langle n\rangle^k$ characterizes  the Poisson distribution. While it may be $g^{(k)}(0)\to 1$ and $g^{(k)}(0)\cdot \langle n\rangle^k \not\to \langle n\rangle^k$, one would have to choose whether assertion $g^{(k)}(0)\approx 1$ or $g^{(k)}(0)\cdot \langle n\rangle^k \approx \langle n\rangle^k$ is the more important one. As a result, we conclude that for this example the characterization of the Poisson behavior by the entropy is in fact better than by the autocorrelation function $g^{(k)}(0)\approx 1$.

\subsection{Construction of the Full Statistics}
\label{subsec:fullrec}
The MEM is not limited to one-dimensional distributions like $p_n$, so one can, e.g., also construct the full statistical distribution $p_{n,N}$ out of moments. For sorting, we define the order of a given moment $\langle n^iN^j\rangle=\sum_{n,N}n^i N^j p_{n,N}$ by $i+j$ visualized in Table~\ref{tab:order}, thus a MED of order $k$ obtains the values of moments $\langle n^iN^j\rangle=\mu_{i,j}, i+j\le k$ as a-priori information. The $k^\text{th}$-order MED can be derived as in the one-dimensional case~\cite{Jaynes57}
\begin{align}
	p^{\text{MED}}_{n,N}=\frac{1}{Z(\lambda)}\exp\left(-\sum_{i+j\le k}\lambda _{i,j}n^iN^j\right)
\end{align}
with the normalization constant $Z(\lambda)= \sum_{n,N}\exp(-\sum_{i+j\le k}\lambda _{i,j}n^iN^j)$. The corresponding Lagrange multipliers are functions of the moment-values $\mu_{i,j}$ and can be calculated by applying the Newton method described in Sec.~\ref{sec:itnewton}.
\begin{table}[h]
\begin{tabular}{ccccccccc}\hline\hline
	{order} &&{photon-carrier moments}\\ \hline
	1 &&  $\langle n\rangle$,\quad $\langle N\rangle$\\
	2 && $\langle n^2\rangle$,\quad $\langle nN\rangle$,\quad $\langle N^2\rangle$\\
	3 && $\langle n^3\rangle$,\quad $\langle n^2N\rangle$ ,\quad $\langle nN^2\rangle$,\quad $\langle N^3\rangle$\\ \hline\hline
\end{tabular}
\caption{ Mixed photon-carrier moments sorted by order.}
\label{tab:order}
\end{table}

\begin{figure}[h]
\vspace{1em}
\includegraphics[width=1\columnwidth]{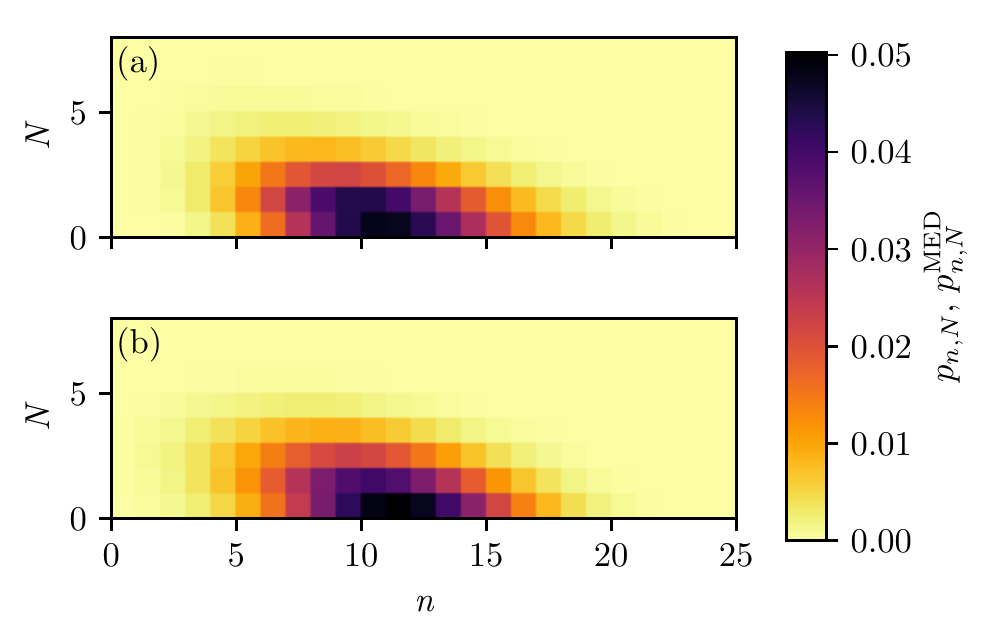}
\vspace{-2em}
\caption{Full statistics MEM-construction for $P = 10$, $\kappa = 1 = \beta$ in second order. (a) The original photon distribution $p_{n,N}$ and (b) the second-order MED approximation $p^{\text{MED}}_{n,N}$.}
\label{fig:fullmed2}
\end{figure}

\begin{figure}[h]
\vspace{1em}
\includegraphics[angle=0,width=0.8\columnwidth]{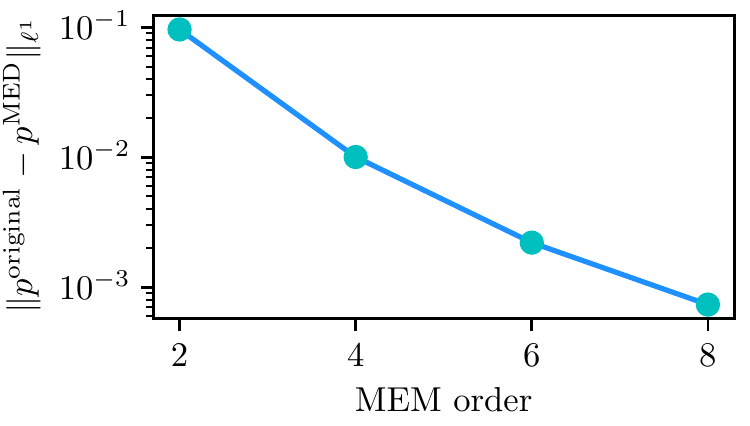}
\vspace{-1em}
\caption{Summed up error of the MED~(\ref{eq:sue}) for even orders as filled circles. Lines are guides to the eyes.}
\label{fig:fullerror}
\end{figure}

After computing the second-order MED for $P = 10$ and $\kappa = 1 = \beta$, one can compare it with the full statistics in Fig.~\ref{fig:fullmed2}. The visible deviation from the proper statistics $p_{n,N}^\text{original}$ is small. In particular, the bell form has been reconstructed very well. In the even higher orders the approximation error falls exponentially as demonstrated in Fig.~\ref{fig:fullerror}, measured by the $\ell^1$-norm
\begin{align}\label{eq:sue}
	\|p^\text{original}-p^\text{MED}\|_{\ell^1}=\sum_{n,N}|p^\text{original}_{n,N}-p^\text{MED}_{n,N}| \ ,
\end{align}
which is the summed-up point-wise absolute distance to the original full statistics $p^\text{original}$. 

Until the ninth-order  the MEM-approximations exist on the approximation space $(0,40)\times(0,60)$. However, only the even-order MEDs should be taken into account, because only their Lagrange multipliers are independent from the choice of the numerical approximation space $\{0,1,\dots, n_\text{max}\}\times \{0,1,\dots, N_\text{max}\}$,   indicating analogously to the one-dimensional case, that, in contrast to the even orders, for the odd orders the MEDs on the full physical range $\mathbb N_0$ do not exist. Furthermore, the general existence conditions for multidimensional MED-cases reveal a more complicated structure than in one dimension~\cite{Junk2000}.

\section{Conclusion}
\label{sec:con}

We have combined equations of motion techniques for nonequilibrium many-particle quantum systems with the maximum entropy method. This extents the range of applicability of the equations of motion techniques significantly as the moments resulting from the steady state can be used to construct unbiased statistics of quantum observables.
After reviewing the maximum entropy method and our physical example, the birth-death model for microcavity lasers with quantum-dot gain, we have explained the mapping from the moments to the Lagrange multipliers by using an iterative Newton method.
Moreover, we have discussed in detail the moment problem, i.e., the fact that not all values of the set of moments can be fitted by a probability distribution.

To confirm the feasibility of our approach we have performed numerical simulations of the birth-death model. From the resulting moments we constructed unbiased photon statistics using the maximum entropy method and compared them to the directly computed photon statistics. Good agreement has been observed which improves when higher-order moments are successively included. The surprising fact that the performance of odd orders is worse than that of the even ones is reported and explained.

We have shown that the zero-crossing of the first Lagrange multiplier signals a qualitative change in the photon statistics.  We have therefore suggested to use this quantity to define the laser threshold. For low spontaneous emission factor~$\beta$ this criterion gives a very similar threshold pump power than the conventional measures. For $\beta$ close to unity, where the conventional measures fail, the first Lagrange multiplier still allows to define an unambiguous threshold.

Also the entropy, which, as the first Lagrange multiplier, comes as a byproduct of the maximum entropy method, provides valuable insight into the system. It shows here that the transition of the photon statistics to a Poisson distribution is much slower than the (higher-order) autocorrelation functions suggest. In fact, for pump powers not too far above the threshold, the photon statistics are better approximated by a Gaussian than by a Poisson distribution.

Finally, we have demonstrated that also the full statistics, including carrier-photon statistics, can be reliably constructed. 

\acknowledgments 
B.~M. acknowledges financial support by the DFG (Project No. WI1986/9-1). We thank S.~Neumeier, T. Lettau, C.~Gies, F.~Jahnke, and P.~Gartner for discussions.


\end{document}